\documentclass[preprint,authoryear,12pt]{elsarticle}

 \usepackage{graphicx}

\usepackage{amssymb}

\journal{Journal of the Mechanics and Physics of Solids}

\begin{document}

\begin{frontmatter}


\author[pt]{Pratyush Tiwary\corref{cor1}}
\ead{pt@caltech.edu}
\cortext[cor1]{Corresponding author}
\address[pt]{Department of Applied Physics and Materials Science, California Institute of Technology, Pasadena, CA 91125, United States}

\author[avdw]{Axel van de Walle}
\ead{avdw@brown.edu}
\address[avdw]{School of Engineering, Brown University, Providence, Rhode Island, Pasadena, RI 02912, United States}

\title{Realistic time-scale fully atomistic simulations of surface nucleation of dislocations in pristine  nanopillars}


\begin{abstract}
We use our recently proposed accelerated dynamics algorithm \citep{mcmd} to calculate temperature and stress dependence of activation free energy for surface nucleation of dislocations in pristine Gold nanopillars under realistic loads. While maintaining fully atomistic resolution, we achieve the fraction of a second time-scale regime. We find that the activation free energy depends significantly on the driving force (stress or strain) and temperature, leading to very high activation entropies. We also perform compression tests on Gold nanopillars for strain rates varying between 7 orders of magnitudes, reaching as low as $10^3/s$. Our calculations show the quantitative effects on the yield point of unrealistic strain-rate Molecular Dynamics calculations: we find that while the failure mechanism for $\langle001\rangle$ compression of Gold nanopillars remains the same across the entire strain-rate range, the elastic limit (defined as stress for nucleation of the first dislocation) depends significantly on the strain-rate. We also propose a new methodology that overcomes some of the limits in our original accelerated dynamics scheme (and accelerated dynamics methods in general). We lay out our methods in sufficient details so as to be used for understanding and predicting deformation mechanism under realistic driving forces for various problems.
\end{abstract}


\end{frontmatter}



\section{Introduction}

Forming a correct picture of dislocation nucleation is central to our understanding of deformation mechanisms at the nano-scale. The initial discoveries by Uchic and subsequent work by various groups \citep{Uchic_science,bin_natmat,nix_acta,greer_afm,volkert,minor,andrew_prl,gao_nanotwin,jenningsacta} have now established that there is a marked increase of yield strength as the specimen size decreases, with significant strain-rate dependence as well. These observations have generally been attributed to the scarcity of dislocation sources (such as Frank-Read sources) in nanosized samples, and having to nucleate dislocations in a perfect crystal (perfect apart from presence of surfaces) \citep{nix_prb,nix_thinfilms,zhu_review,weinberger_jmc}. As such there have been numerous attempts to link simulations of dislocation nucleation processes to experimentally observed mechanical behavior - in fact a lot of crucial insight has come from simulations \citep{bulatov_nature,bulatov_naturemat,moriarty,juli_nature,abraham_pnas,mao_nanowire,juli_prl,weicai_pnas}. Nanoindentation experiments \citep{nanoind_science,schuh_nature,schuh_mattoday}, Scanning Electron Microscopy combined with Nanoindentation \citep{sementor}, and High Resolution Transmission Electron Microscopy (HRTEM)\citep{sub10nm} are now sufficiently advanced for one to hope for a direct match between simulations and experiments \citep{zhu_prb,mao_acta,greer_matscea,sub10nm}. \newline

There are three broad classes of techniques for such simulations, often used in conjunction with each other and along with approaches such as Transition State Theory: classical Molecular Dynamics, continuum based methods and \emph{ab initio} techniques. \emph{Ab initio} simulations, though they often provide insight into mechanical behavior \citep{arias_prl,sitch_prb,woodward_prl2,woodward_prl1,carter1}, are still restricted to very small sizes, less than a hundred atoms typically (although there are several promising attempts at bridging this length-scale gap for \emph{ab initio} calculations \citep{carter2,gavini1,gavini2,trinkle_green}). The achievable time-scales are also typically restricted to less than a few picoseconds. Continuum based methods are another elegant option capable of dealing with a variety of length and time scales, though they suffer from not providing atomic scale resolution and assuming elastic behavior even at dislocation cores 
\citep{hirthe,bulatov_cont,weinberger_jmps}. Classical Molecular Dynamics (MD) can be helpful in gaining quantitative insight into mechanical behavior at various length scales (nanometers to microns or larger) \citep{zhou_science,bulatov_cms,diao,park_jmps,srolovitz,gao_nanotwin}. MD does not assume much apart from the form of interatomic interaction, which is typically developed by fitting to first principles or experimental data. The availability of quality interatomic force-fields \citep{baskes,bulatov_pot,grochola,srolovitz} and increase in computer power has led to a tremendous increase in the popularity of MD over the last decade.\newline

However, most of the interesting dynamics happens only as the system moves from one energy basin to another through infrequent, rare events. Most of the simulation time gets spent with the system staying stuck in some energy basin \citep{metadynamics}. This behavior, combined with the femtosecond timestep required for total energy conservation, gives rise to a major limitation of MD: the time-scale problem \citep{voter_review,ejp_yip,diffusivemd}. Even with the advent of powerful super-computers, MD simulations are unable to reach more than a few nanoseconds of time if the system size is more than a few hundred atoms. Thus while laboratory strain-rates are typically in the range $10^{-5}$-$10^3$/sec, with corresponding activation free energies being around 30$k_BT$, MD is unable to go slower than $10^7$/sec strain-rate, corresponding to free energies of around 5$k_BT$ or lower\citep{juli_prl,asbhd,weicai_pnas,mcmd}. One approach to get around this shortcoming is to perform 0 temperature Nudged Elastic Band calculation \citep{neb} of the activation free energy and how it varies with applied stress, and then either assume it to be temperature independent, or assume a phenomenological model for its variation with temperature (such as multiplying it with an empirical temperature dependent scaling factor) \citep{weinberger_jmps}. These approaches can sometimes work well, but  as shown by \citet{asbhd,weicai_pnas} and in this current work, can sometimes lead  to significant inaccuracies in the predictions, such as errors of several orders of magnitude in the nucleation rate, or even qualitatively incorrect phase transitions\citep{curtin_actamat}.\newline

We recently proposed a hybrid method that combines the strengths of MD and Monte Carlo (MC) simulations in an easily implementable manner and enables one to reach milliseconds and longer times for several thousands of atoms \citep{mcmd}. The method was recently applied to vacancy diffusion in BCC Fe at low temperature as well as calculation of stress-strain behavior for Au nanowire compression. In both cases it was found to predict correct dynamics, with excellent scaling in computational efficiency with system size. The algorithm was particularly designed to be used on massively parallel computer systems. The main advantages of this new method over existing accelerated MD techniques (see, for e.g., \cite{voter_prl,voter_review,fichthorn,kappa}) are that (i) it provides a statistically more accurate ``real'' time scale (which is important when determining the actual strain rate in a simulation and time-dependent forces in general), (ii) it does not rely on harmonic transition state theory (which is important when the object of interest is the entropy of activation or migration), (iii) it does not require the specification of the degrees of freedom of interest, which is crucial when the mechanisms are complex and involve the movement of many atoms and (iv) the efficiency of estimating the ``real'' time scale (as in (i)) improves linearly with number of computer processors employed for the calculation. We would like to point out that methods like Hyperdynamics, at least in their originally proposed flavor, do not have the limitation of specifying degrees of freedom. 
\newline

It has remained an unsolved problem so far to design and peform fully atomistic simulations that could provide a picture of temperature dependent activation free energies of dislocation nucleation from surfaces at realistic loads and loading rates. Such a picture is key to linking experimental results with simulation predictions \citep{curtin_actamat,weicai_pnas,weinberger_jmps}. The critical nucleus for surface nucleation can be as small as a few atomic planes, thus questioning the applicability of continuum methods. As for classical MD, the time-scale achievable is several orders of magnitude smaller than experiments, thus limiting MD simulations to regimes of extremely high nucleation rates. With our recently proposed hybrid MC-MD method that allows us to achieve extended time-scales while still maintaining atomistic resolution, we are able to study the temperature dependence of activation parameters for surface nucleation of dislocations in pristine nanowires and obtain several significant results in an activation regime actually achievable in laboratory experiments. The specific problem we consider pertains to several nano-indentation experiments where it was found that even if the applied stress on a sample is in the elastic regime, yielding could occur after a certain statistically distributed waiting time \citep{ngan_prl,ngan_philmag,schuh_prb}. We perform fully atomistic simulations of this time-dependent incipient plasticity behavior in Gold nanowires, reaching hundreds of milliseconds time-scales for several thousand atoms. After collecting statistics for various temperatures and applied stresses, we then derive the full picture of stress and temperature dependence of the activation free energy. \newline

In Section 2, we provide a brief summary of out hybrid method in order to facilitate further discussion and keep this paper self-contained. For a more detailed explanation of terms we refer the reader to  \citet{mcmd}. In the current paper, we also propose a new adiabatic switching technique that significantly reduces the number of input parameters in our hybrid MC-MD approach and eliminates some of the fundamental limitations of our earlier implementation (that were shared by related algorithms \citep{voter_prl}). The algorithms employed here make it possible to achieve linear scaling in efficiency of estimating the accelerated time as the number of parallel processors employed is increased. We describe our algorithm and its implementation in sufficient detail for researchers to be able to use it for their problem of interest, and hope that it will be found helpful for modeling a variety of mechanical behavior problems.\newline

\section{Details of calculations}

\subsection{Choice of interatomic potential}
There are several good potentials available for modeling mechanical behavior of Gold \citep{baskes,caiye,foiles,grochola}. The embedded atom method potential by \citet{grochola} gives very realistic values for the surface energy and the stacking fault energy \citep{srolovitz}: the stacking fault energy from the potential by \citet{grochola} is 42 mJ/$m^2$ while the experimental value for it is in the range 32-46 mJ/$m^2$. Since the current paper deals with nucleation of dislocations from surfaces, we choose the Grochola potential. This potential was also used and found to perform very well in a recently published joint computational and experimental work studying dislocation behavior in sub-10nm Gold nanowires \citep{sub10nm}. \citet{srolovitz} provide a critical comparison of this potential with other available potentials for various physical properties relevant to the current work.

\subsection{Hybrid stochastic and deterministic technique for achieving realistic time-scales}
\subsubsection{Summary of ideas}
We recently proposed using a combination of MD and MC techniques for achieving long time scales \citep{mcmd}. Our approach is built upon minimizing the MD time spent in low-lying energy basins, and instead using 2 kinds of MC simulations. One (a) seeks to properly thermalize the system between infrequent events, thereby minimizing artifical correlations, and the other (b) provides independent control over the accuracy of the time-scale correction. When the potential energy $V(x)$ of the system (where $x$ is a point in the 3-N dimensional configuration space for a system with N particles) is above a certain $V_0$, the system evolves as per regular MD (see Fig. 1 in \citet{mcmd}). This high energy region of the phase space is the one containing the interesting but infrequent events.
When the system potential energy goes below $V_0$, we allow MD to continue until the system has lost memory of how it entered this well (defined as all points $x$ such that $V(x) \leq V_0$). We found that a simple and appropriate criterion to check for this memory loss is when the energy reaches the system's mean energy at that temperature (although other choices are possible, such as letting MD continue for a sufficiently long, user-specified, time or for a random length of time drawn from a user-specified exponential distribution).
During this thermalization time, the system may escape the well, in which case the system simply keeps evolving via MD. Most likely, however, the system will not escape the well during that time. When this happens, we stop MD and launch the first MC simulation (called MC $a$). \newline

MC $a$ runs with a perfectly uniform potential inside the well, rejecting all moves that lead to $V(x)>V_0$. The purpose of MC simulation $a$ is to generate a new, properly thermalized, starting point for MD. MC $a$ serves to de-correlate the system between the time it entered the well and when it leaves it. This is necessitated by the rare event hypothesis - on an average, the system should have lost memory of how it entered the well before leaving it again. MD resumes with positions drawn from the last MC $a$ step that visited the boundary of the well(defined rigorously in Eq.(\ref{eq:lid})), and velocities drawn from a Maxwell-Boltzmann distribution in the half space pointing outward of the well. $V_0$ can be picked high or low depending on the speed-up relative to MD we seek for a particular application. The method is formally correct for any choice of $V_0$; a higher choice of $V_0$ limits our ability to monitor the detailed dynamics of some events. \newline

We also need to estimate the expected value of the time the system would have spent in the energy well \textbf{W}, which can be calculated as the reciprocal of the flux exiting the well:
\begin{equation}
\label{eq:time}
t_\textbf{W} = \lim_{w \rightarrow 0} ( \langle  {\overline{v}\over w} \; 1(x \in \textbf{S$_w$}) \rangle )^{-1}
\end{equation}
where the average $\langle\cdots\rangle$ is taken over $x$ drawn from the well \textbf{W} with a probability density proportional to $e^{-V(x)/(k_B T)}$. $k_B$ is Boltzmann's constant, T is the temperature and the following definitions hold:
$1(A)$ equals 1 if the event $A$ is true and 0 otherwise,
\textbf{S$_w$} is a shell of width $w$ at the boundary of the well \textbf{W}, which can be defined in the limit of small $w$ as
\begin{equation}
\label{eq:lid}
\textbf{S$_w$} = \{x: |V(x)-V_0| \leq w|\nabla V(x)|/2 \}
\end{equation}
$\overline{v}$ denotes the mean projection of a Maxwell-Boltzmann-distributed velocity along the unit vector $u$ parallel to $\nabla V(x)$, conditional on $v \cdot u>0$. When all atoms have the same mass $m$, $\overline{v} =\sqrt{k_B T/{2 \pi m}}$ (a general expression can be found in \cite{mcmd}). \newline 

Since Eq.(\ref{eq:time}) involves an average, it can be approximated using MC simulations. We make use of the system's ergodicity, replacing the time average (that would require us to wait for long times for it to converge) by an ensemble average. Thus in parallel to MC $a$, we launch several instances (as many as number of available processors) of a second kind of MC simulation, called MC $b$, to estimate the time-scale correction. A most straightforward implementation of this still won't be as effective in estimating the average in Eq.(\ref{eq:time}) because the shell \textbf{S$_w$} would be visited very rarely. Thus, to improve the efficiency in estimating Eq.(\ref{eq:time}), we proposed in \citet{mcmd} using a biased potential $V^* (x)$, which is the same as the true potential $V(x)$ in the high energy regions but lifted up in the energy basins \citep{voter_prl,voter_review}. Several lifting (or biasing) schemes are available for use in this \citep{voter_review}. A simple importance sampling expression (as detailed in \citet{mcmd}) can thus give us the following time-correction:

\begin{equation}
\label{eq:t_bias}
t_\textbf{W} = \lim_{w \rightarrow 0} \frac{  \langle  e^{-{{\beta}}(V(x)-V^{*}(x))} \rangle^* }
{\langle {\overline{v}\over w} {e^{-{{\beta}}(V(x)-V^{*}(x))} }1(x \in \textbf{S$_w$}) \rangle^* }
\end{equation}

where $\langle \cdots \rangle*$ denote expectations taken under a density proportional to $e^{-\beta V^*(x)}$, in which $\beta$ is $1/(k_B T)$. This approach works well, but there is a fundamental trade-off that limits its usefulness. Lifting the biased potential more and more leads to the energy shell \textbf{S$_w$} being visited more frequently and should lead to greater computational efficiency in estimating Eq.(\ref{eq:time}). However a compensating effect leads to a decrease in efficiency beyond a certain amount of biasing. This is because increased biasing of the potential leads to noisier statistical everaging of the time in Eqs.(\ref{eq:time}) and (\ref{eq:t_bias}) -  as biasing increases, $(V(x)-V^{*}(x))$ becomes a large number causing $e^{-\beta(V(x)-V^{*}(x))}$ to dramatically increase. This point has been discussed in detail in \citet{fichthorn,asbhd,mcmd}. This trade-off is a general problem that arises in importance sampling methods, where one wishes to pick the sampling scheme that leads to maximum variance reduction \citep{rubino}. 

\subsubsection{Adiabatic switching technique}
We now propose a technique that bears some resemblance to adiabatic switching methods (see, e.g. \citep{tuckerman}) that helps us deal with the trade-off discussed above, and also eliminates the need for picking a particular biasing scheme. The motivation here is to avoid the statistical noise in Eq.(\ref{eq:t_bias}) that arises as the biased potential $V^{*}(x)$ becomes increasingly different from the true potential $V(x)$. To avoid this noise in sampling, the system is continuously, \emph{adiabatically} switched from $V\left( x\right) $ (the true potential) to $V_{0}$ (a flat potential within the well, identical to the potential used in MC simulation $a$ for thermalization). We now formally derive the method.\newline

Let $\hat{V}\left( x,\alpha \right) $ smoothly interpolate between $\hat{V}%
\left( x,0\right) \equiv V\left( x\right) $ and $\hat{V}%
\left( x,1\right) \equiv V_{0}$. Then we can express the ensemble average in Eq.(\ref{eq:time}) as below (working in terms of $rate=1/time$):%
\begin{eqnarray}
\label{eq:adb1}
rate &=&\lim_{w \rightarrow 0}\frac{\int {\bar{v}\over{w}} 1\left( x\in S_{w}\right)
e^{-\beta \hat{V}\left( x,0\right) }dx}{\int e^{-\beta \hat{V}\left(
x,0\right) }dx} \nonumber \\
&=&\lim_{w \rightarrow 0}\frac{\int  {\bar{v}\over{w}}  1\left( x\in S_{w}\right) e^{-\beta \left( 
\hat{V}\left( x,0\right) -\hat{V}\left( x,1\right) \right) }e^{-\beta \hat{V}%
\left( x,1\right) }dx}{\int e^{-\beta \hat{V}\left( x,1\right) }dx}\left( 
\frac{\int e^{-\beta \hat{V}\left( x,1\right) }dx}{\int e^{-\beta \hat{%
V}\left( x,0\right) }dx}\right)  \nonumber \\
&\equiv &\lim_{w \rightarrow 0} \left\langle {\bar{v} 1\left( x\in S_{w}\right)\over{w}} e^{-\beta
\left( \hat{V}\left( x,0\right) -\hat{V}\left( x,1\right) \right)
}\right\rangle _{1}R
\end{eqnarray}%
where $dx$ denotes a differential volume in 3-N dimensional configuration space for N particles, the integration being performed over entire configuration space within the well \textbf{W} and the expected value $\left\langle \cdots \right\rangle _{\alpha }$ in Eq.(\ref{eq:adb1})  is defined by
\begin{equation}
\label{eq:adb2}
\left\langle \cdots \right\rangle _{\alpha }=\frac{\int\left( \cdots
\right) e^{-\beta \hat{V}\left( x,\alpha \right) }dx}{\int e^{-\beta \hat{%
V}\left( x,\alpha \right) }dx}
\end{equation}
Below we define the term $R$ in Eq.(\ref{eq:adb1}) and re-express it in a computationally tractable form:
\begin{eqnarray}
\label{eq:adb3}
R &=&\frac{\int e^{-\beta \hat{V}\left( x,1\right) }dx}{\int e^{-\beta 
\hat{V}\left( x,0\right) }dx} \nonumber \\
&=&\exp \left( \ln \int e^{-\beta \hat{V}\left( x,1\right) }dx-\ln
\int e^{-\beta \hat{V}\left( x,0\right) }dx\right)  \nonumber \\
&=&\exp \left( \int_{0}^{1}\left( \frac{\partial }{\partial \alpha }\ln
\int e^{-\beta \hat{V}\left( x,\alpha \right) }dx\right) d\alpha \right) 
\nonumber \\
&=&\exp \left( -\beta \int_{0}^{1}\frac{\int \frac{\partial \hat{V}\left(
x,\alpha \right) }{\partial \alpha }e^{-\beta \hat{V}\left( x,\alpha \right)
}dx}{\int e^{-\beta \hat{V}\left( x,\alpha \right) }dx}d\alpha \right) \nonumber  \\
&=&\exp \left( -\beta \int_{0}^{1}\left\langle \frac{\partial \hat{V}\left(
x,\alpha \right) }{\partial \alpha }\right\rangle _{\alpha }d\alpha \right) 
\end{eqnarray}
With this we can now write the rate in Eq.(\ref{eq:adb1}) as
\begin{equation}
\label{eq:adb4}
rate =  \lim_{w \rightarrow 0}\bar{v} \left\langle  {{1\left( x\in S_{w}\right)}\over{w}} e^{-\beta\left( \hat{V}\left( x,0\right) -\hat{V}\left( x,1\right) \right)}\right\rangle _{1}\exp \left( -\beta \int_{0}^{1}\left\langle \frac{\partial \hat{V}\left(
x,\alpha \right) }{\partial \alpha }\right\rangle _{\alpha }d\alpha \right) 
\end{equation}

\begin{figure}[htp]
\centering
 \includegraphics[width=55mm]{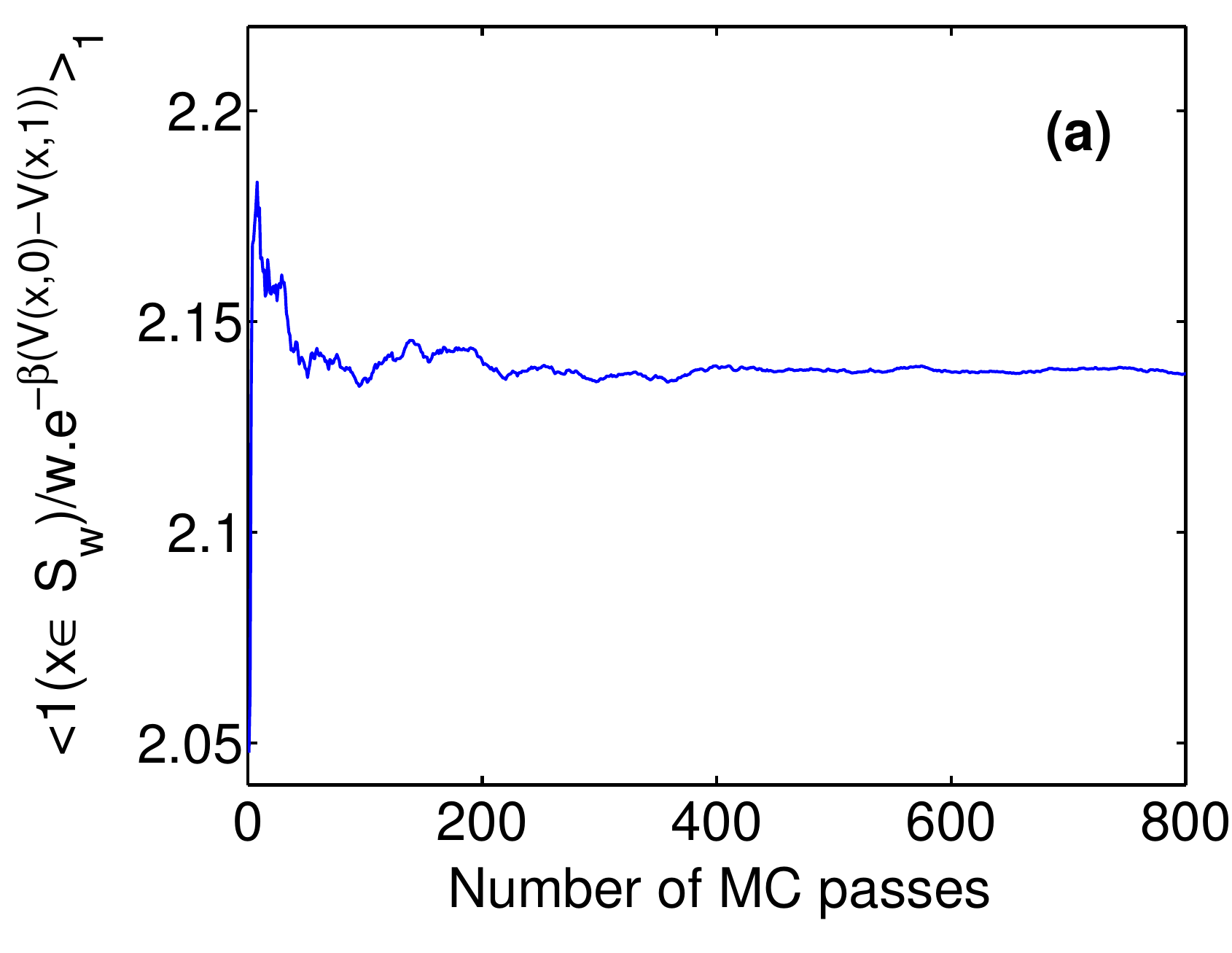}
 \includegraphics[width=55mm]{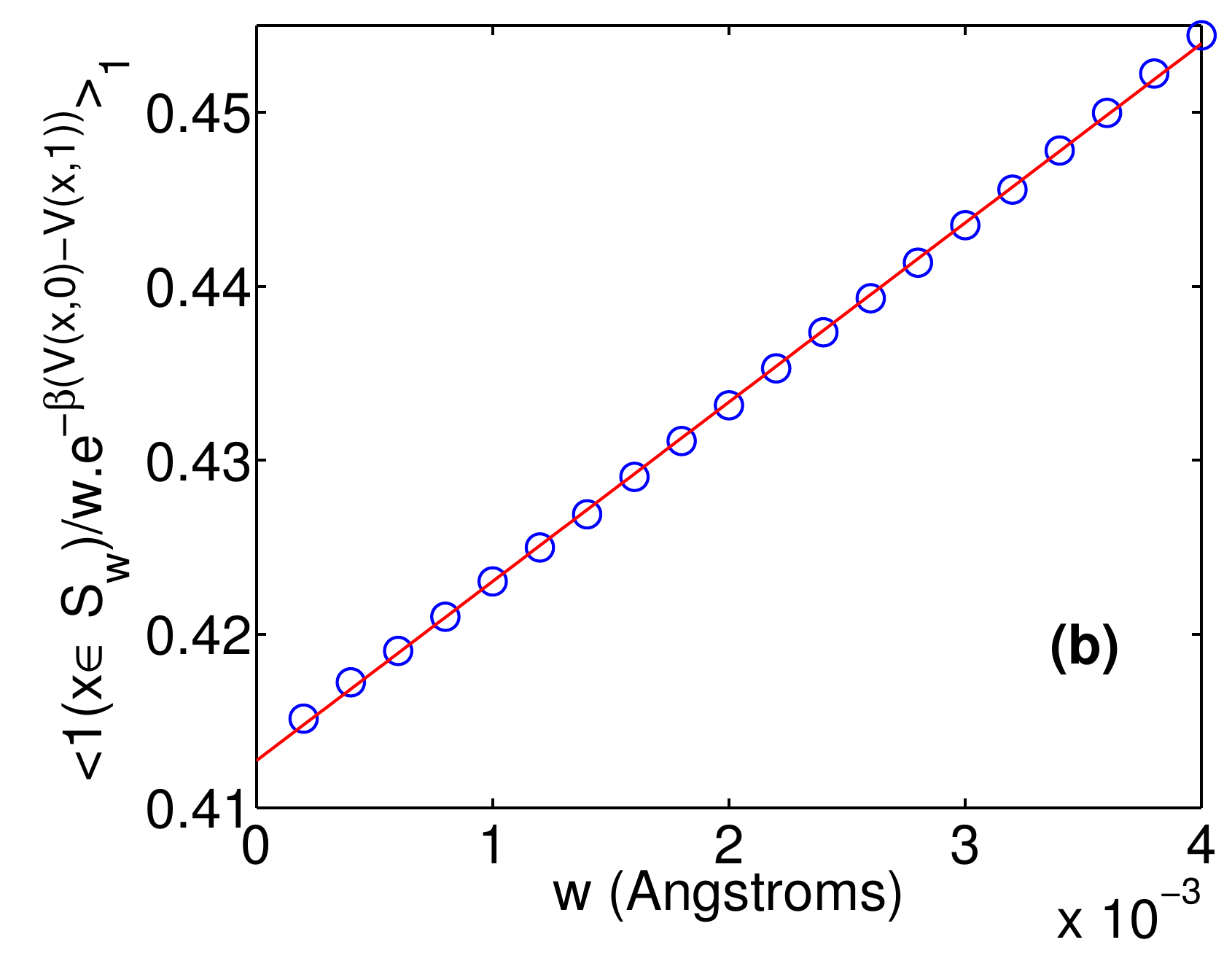}
 \caption[1]
{(a) The 2nd part in Eq.(\ref{eq:adb4}), i.e. $\left\langle{{1\left( x\in S_{w}\right)}\over{w}} e^{-\beta\left( \hat{V}\left( x,0\right) -\hat{V}\left( x,1\right) \right)}\right\rangle _{1}$, can be evaluated in a very small number of MC passes as explained in the text. (b) Calculating $\lim_{w \rightarrow 0}\left\langle{{1\left( x\in S_{w}\right)}\over{w}} e^{-\beta\left( \hat{V}\left( x,0\right) -\hat{V}\left( x,1\right) \right)}\right\rangle _{1}$ using linear extrapolation. Calculations are for a Au nanowire with 2016 atoms, 2.5nm in diameter and 7.5nm in height, at 300 K.}
\label{fig:flat}
\end{figure}

We now make a few observations regarding the above expression. It involves 3 independent parts. The first is $\bar{v}$, which we already know as $\sqrt{k_B T/{2 \pi m}}$ for identical atoms. We could keep $\bar{v}$ inside the ensemble average to cover the general case of unequal masses in which $\bar{v}$ may depend on $x$. The second part in Eq.(\ref{eq:adb4}) is $\lim_{w \rightarrow 0}\left\langle{{1\left( x\in S_{w}\right)}\over{w}} e^{-\beta\left( \hat{V}\left( x,0\right) -\hat{V}\left( x,1\right) \right)}\right\rangle _{1}$. This is non-0 only when $x\in S_{w}$, and whenever it is non-0, the difference $\hat{V}\left( x,0\right) -\hat{V}\left( x,1\right)$ is a very small number (see Eq.(\ref{eq:lid}))). Since this average is calculated with a flat potential $\hat{V}\left( x,1\right)$, the boundary $x\in S_{w}$ is visited frequently, and thus the second term in Eq.(\ref{eq:adb4}) can be evaluated very quickly - in a few MC passes as shown in  figure \ref{fig:flat}(a). We calculate $M \equiv \left\langle{{1\left( x\in S_{w}\right)}\over{w}} e^{-\beta\left( \hat{V}\left( x,0\right) -\hat{V}\left( x,1\right) \right)}\right\rangle _{1}$ for a few values of $w$, and simple linear extrapolation gives the desired limit, as shown in figure \ref{fig:flat}(b). The third part in Eq.(\ref{eq:adb4}) is $exp \left( -\beta \int_{0}^{1}\left\langle \frac{\partial \hat{V}\left(
x,\alpha \right) }{\partial \alpha }\right\rangle _{\alpha }d\alpha \right)$. Here, the average $\langle \frac{\partial \hat{V}(x,\alpha) }{\partial \alpha }\rangle _{\alpha }$ does not contain any exponentials, and thus no terms that could blow-up and lead to noisy estimates and slow convergence. \newline

We now need to pick up a switching scheme for $\hat{V}(x,\alpha)$, i.e. an interpolation scheme between $\hat{V}(x,0)$ and $\hat{V}(x,1)$. We picked the simplest scheme - a linear switching model - and found it to work very well:

\begin{equation}
\label{eq:switch}
\hat{V}(x,\alpha) = (1-\alpha)V(x) + \alpha V_0
\end{equation}

With this, Eq.(\ref{eq:adb4}) for the rate of escaping energy basins bounded by $V(x)<V_0$ becomes
\begin{equation}
\label{eq:adb5}
rate = \lim_{w \rightarrow 0} {\bar{v}}\left\langle{{1\left( x\in S_{w}\right)}\over{w}} e^{-\beta\left( \hat{V}\left( x,0\right) -\hat{V}\left( x,1\right) \right)}\right\rangle _{1}\exp \left( \beta \int_{0}^{1}\langle 
{ V(x)-V_0}\rangle _{\alpha }d\alpha \right) 
\end{equation}

\begin{figure}[htp]
\centering
 \includegraphics[width=60mm]{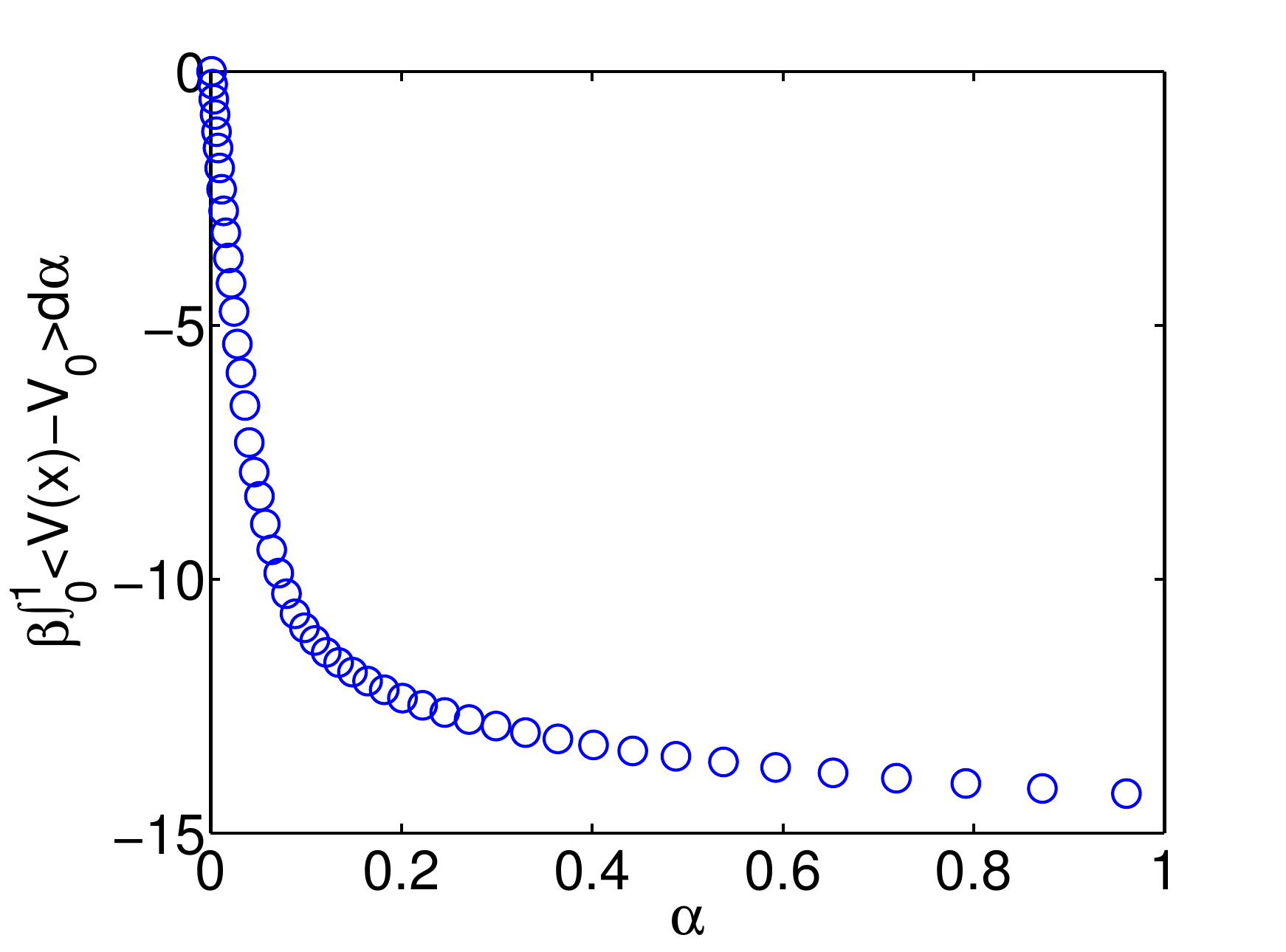}
 \includegraphics[width=60mm]{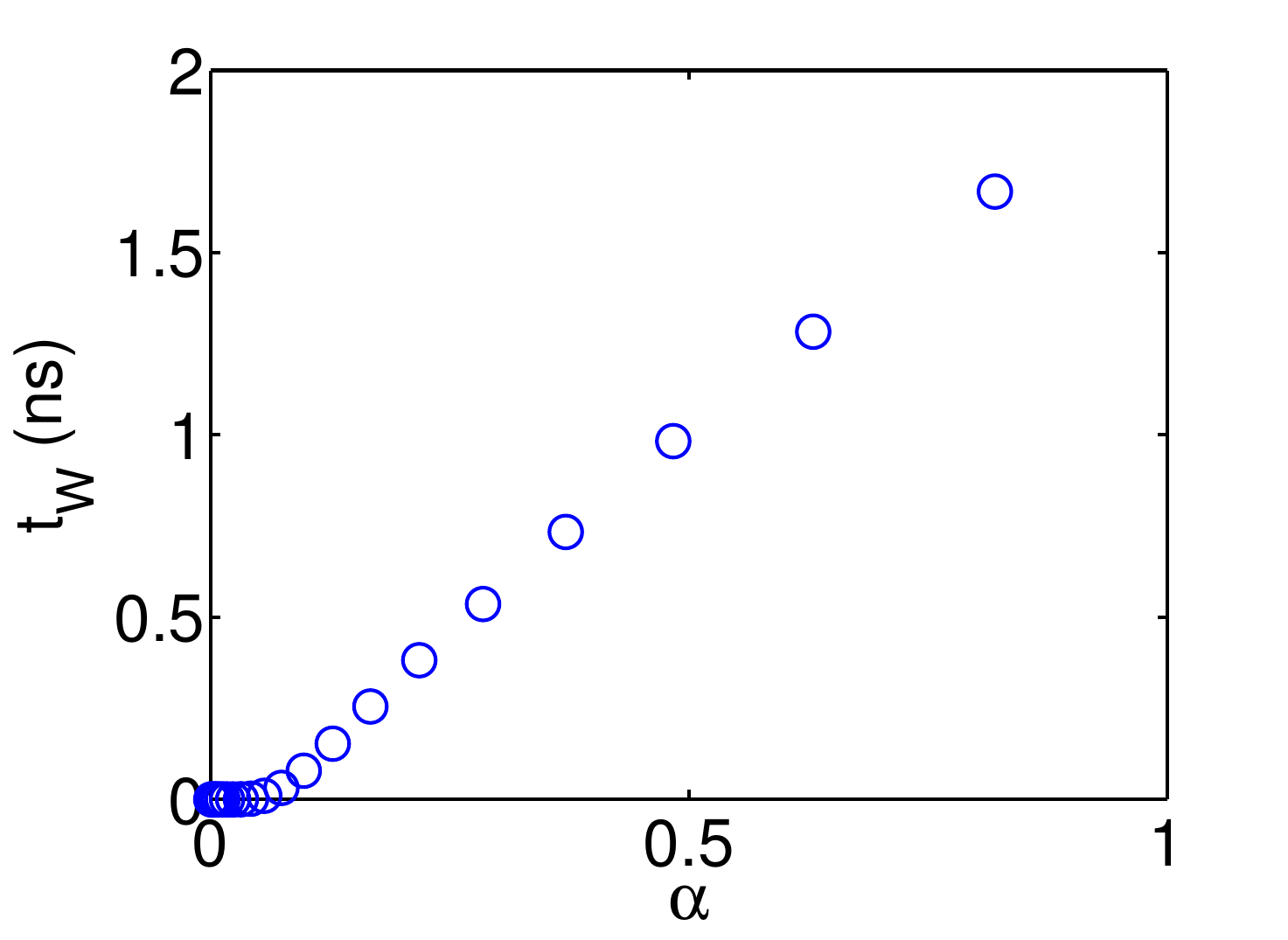}
 \caption[1]
{(a) Change in  $ \left( -\beta \int_{0}^{1}\left\langle \frac{\partial \hat{V}\left(
x,\alpha \right) }{\partial \alpha }\right\rangle _{\alpha }d\alpha \right)$ as a function of $\alpha$ as the simulation progresses (see Eq.(\ref{eq:adb4})). (b) Expected value of the time spent in the energy well \textbf{W}. Calculations are for a Au nanowire with 2016 atoms, 2.5nm in diameter and 7.5nm in height, at 300 K. }
\label{fig:adb}
\end{figure}

Thus to summarize till this point, to calculate Eq.(\ref{eq:adb5}), we first do a quick MC simulation using a flat potential to get $\lim_{w \rightarrow 0}\left\langle{{1\left( x\in S_{w}\right)}\over{w}} e^{-\beta\left( \hat{V}\left( x,0\right) -\hat{V}\left( x,1\right) \right)}\right\rangle _{1}$, as shown in figure \ref{fig:flat}. We then vary $\alpha$ adiabatically during the simulation, going from $\alpha=0$ to $\alpha=1$. We perform a series of MC simulations, with the Hamiltonian of the system evolving as per Eq.(\ref{eq:switch}) as the simulation time progresses. A typical evaluation of Eq.(\ref{eq:adb5}) done as per this scheme is shown in figure \ref{fig:adb}, where we show the change in the following two as a function of $\alpha$: (a)  $ \left( -\beta \int_{0}^{1}\left\langle \frac{\partial \hat{V}\left(
x,\alpha \right) }{\partial \alpha }\right\rangle _{\alpha }d\alpha \right)$, related to the 3rd part in Eq.(\ref{eq:adb5}), and (b) the expected value of the time spent in the energy well \textbf{W}.

It may be useful to emphasize one point: The reason we do not immediately start MC as soon as $V(x)$ falls below $V_0$ is because the rate expression (Equation (\ref{eq:time})) is only valid conditional on the system being initialized at a Boltzman-distributed random position within the well. If we start MC at the boundary of the well, this assumption is violated. Although it may seem that, by running MD within the well for some time before calculating the escape rate, we slightly overestimate the time spent in the well, this is not the case. The distribution of escape time from the well is independent of the time already spent in the well (since it follows an exponential distribution). Another way to see that there is no time over-counting is to observe that, during this MD trajectory in the well, there is also a small probability that the system escapes the well, so we are not artificially constraining the system to remain in the well for a longer time. The above scheme is similar to what is done in the parallel replica method \citep{parrep,parrep_driven}. It provides a convenient way to deal with so-called re-crossing events that typically affect the accuracy of transition-theory based estimates of escape times method \citep{voter_review,kappa}.

\subsection{Simulation set-up and compression testing}

\begin{figure}[htp]
\begin{center}
 \includegraphics[width=50mm]{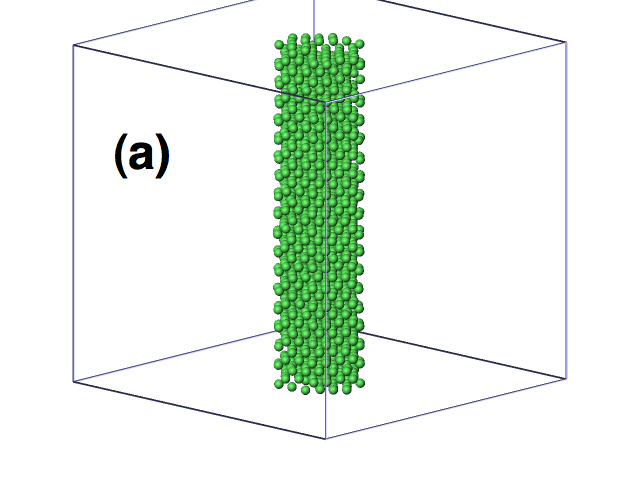}
 \includegraphics[width=50mm]{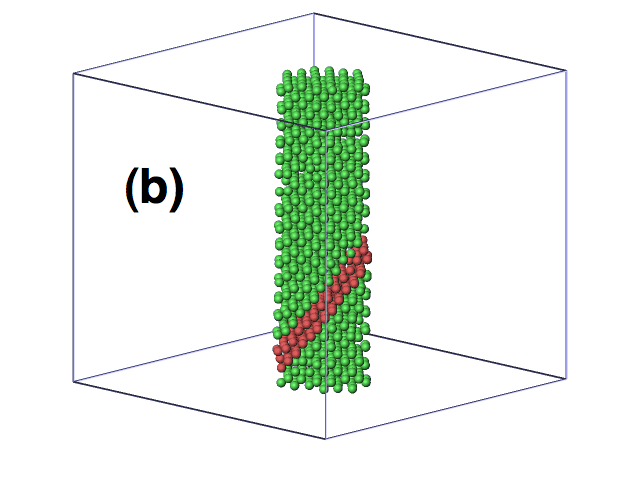}
 \includegraphics[width=50mm]{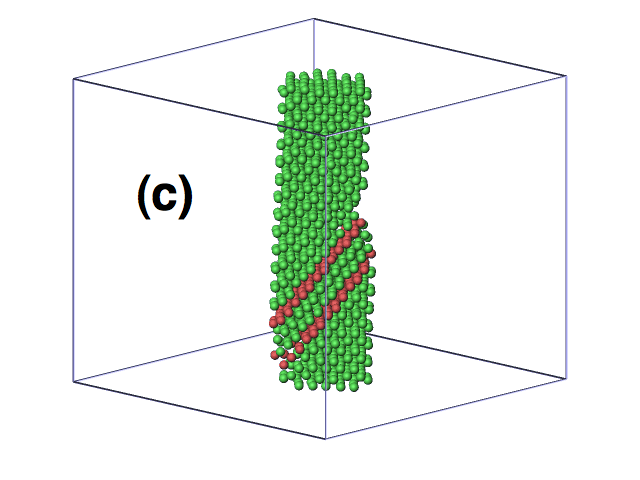}
 \end{center} 
 \caption[1]
{Perspective view of the nanopillar (a) before compression and (b) at various stages of compression when multiple partial dislocations have nucleated. The periodic supercell is also shown. Coloring is as per Common Neighbor Analysis\citep{cna}, where green denotes FCC, red denotes HCP. The surface atoms (identified as atoms that are neither FCC nor HCP) have been removed to bring the slip plane into clarity. Visualization was carried using the package OVITO\citep{ovito}. Movie in supplementary information shows the deformation process between these snapshots at a strain-rate of 1000/sec.}
\label{fig:slip}
\end{figure}

\begin{figure}[htp]
\begin{center}
 \includegraphics[width=100mm]{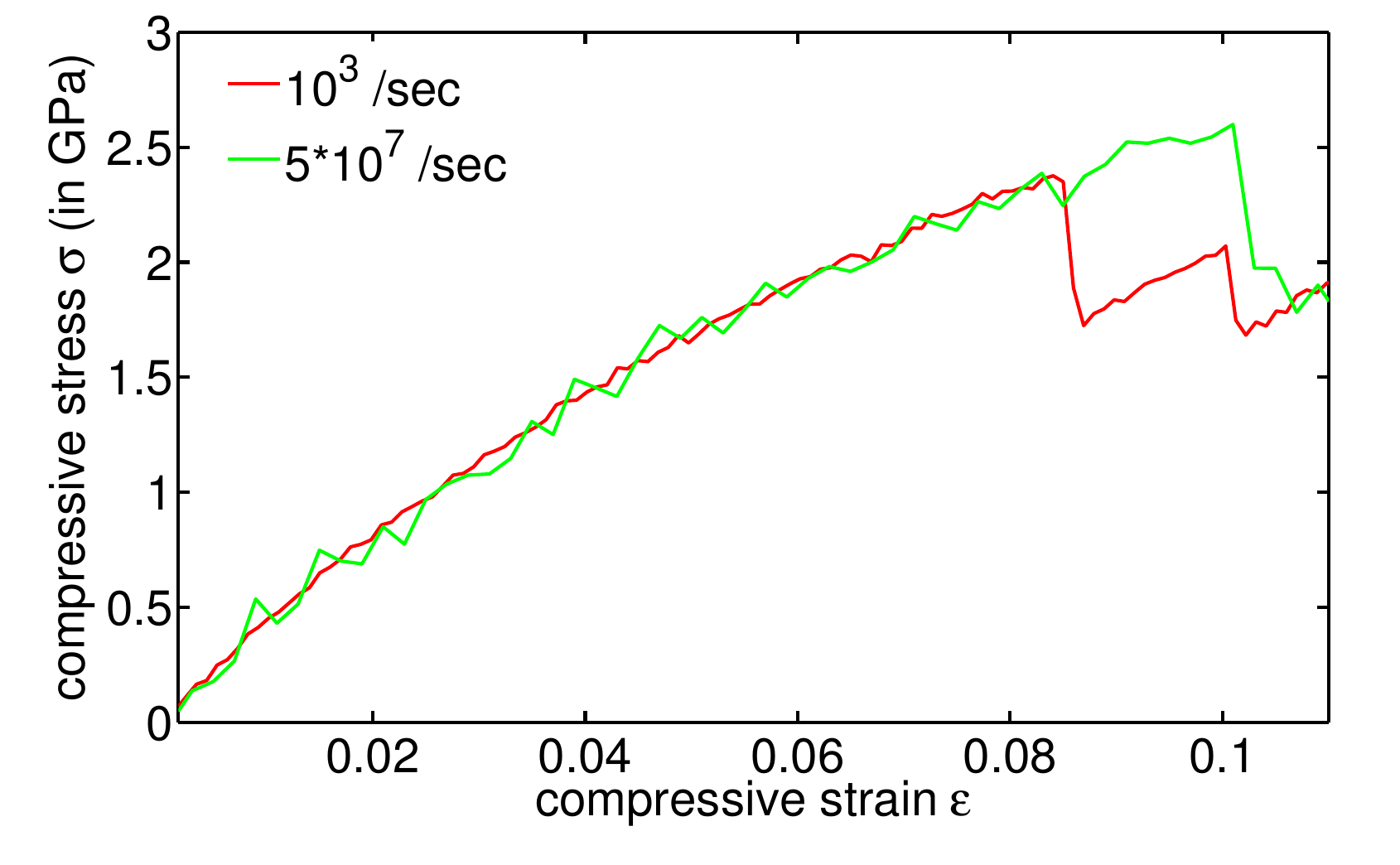}
 \end{center} 
 \caption[2]
{The stress versus nominal strain curve of Au nanowire under $\langle 001 \rangle$ compression (with inverted sign of stress). The data represents stress relative to surface stress at 0 strain as explained in text. The stress-strain plots are shown for two different strain rates. The green line denotes our calculations for a strain rate of 5x$10^7$/sec, which is a commonly used strain rate in current-day MD. The red line shows our calculations for a strain rate of $10^3$/sec.}
\label{fig:stressstrain}
\end{figure}

We first report the stress-strain plots for $\langle 001 \rangle$ compression of pristine cylindrical Au nanowires. The cylinder was initially carved out from perfect FCC lattice and before compression, it was 2.5nm in diameter and 7.5nm in height, comprising 2016 atoms (see figure \ref{fig:slip}) with periodic boundary conditions imposed along all three directions. The cylinder axis $z$ is also the compression axis $\langle001\rangle$. Thus all sites along the length of the wire are now equivalent sites for nucleation. For the other two directions, we do not strictly need periodic boundary conditions, but we nevertheless apply it for computational ease. The dimensions of the supercell in the \emph{x} and \emph{y} directions are both around 75$\AA$, which is much larger than the range of the EAM potential employed (5.5$\AA$)\citep{grochola}. As such there is no artifact from the pillar interacting with its images in these two directions. The cylinder was first equilibrated for 500 ps before beginning the compression, which was carried out by uniformly re-scaling the z-coordinates of all atoms. The atomic virial stress was used to obtain the Cauchy stress\citep{srolovitz}. The stress at zero nominal strain is non-zero and tensile, and arises from the surface stress (see \citet{srolovitz, weinberger_jmps} for a more detailed explanation). We adjust for this, and as such figure \ref{fig:stressstrain} provides the stress span, i.e. the stress at a strain $\epsilon$ relative to the stress at 0 strain.  We present the resulting stress($\sigma$) versus nominal strain($\epsilon$) plots for 2 different strain rates $\dot{\epsilon}$: 5x$10^7$/sec, a strain rate value used in current day state-of-the-art MD simulations, and $10^3$/sec. To the best of our knowledge, the latter is a strain rate several orders of magnitude slower than any reported calculation for a nanowire, and is a value much closer to that can actually be achieved in laboratory experiments on nanowires by using state-of-the-art nanoindentors and steering hardware (Robert Maass, private communication). \newline

\begin{figure}[htp]
\begin{center}
 \includegraphics[width=80mm]{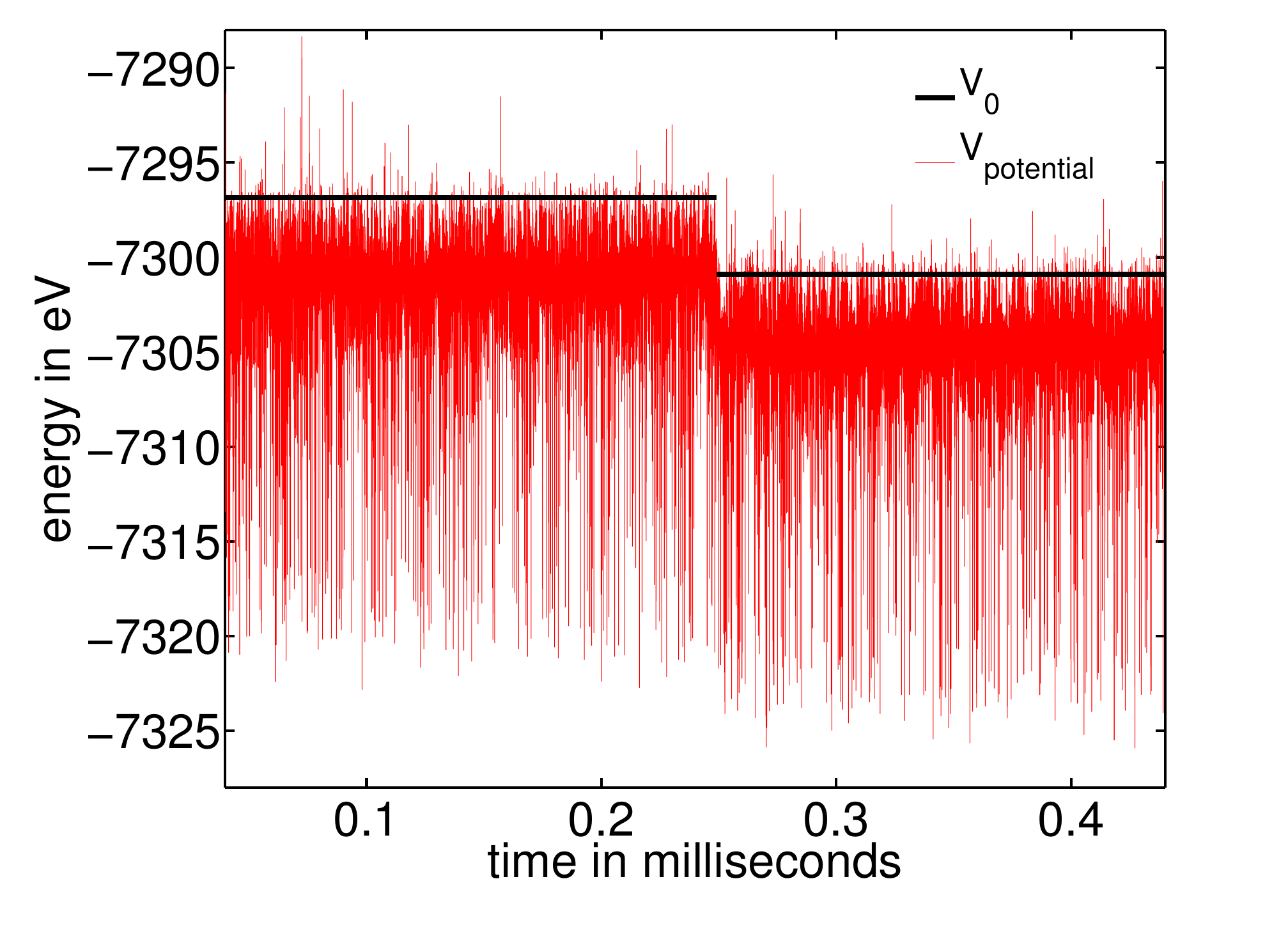}
 \end{center} 
 \caption[1]
{Typical variation of the energy lid $V_0$ (black line) and potential energy (red line) as a function of time. $V_0$ is adjusted as per the change in mean potential energy every few thousand MD steps. A sharp drop in the potential energy and accordingly in $V_0$ indicates that a nucleation event has occurred.}
\label{fig:energyprofile}
\end{figure}

For the strain rate of 5x$10^7$/sec, it was sufficient to perform ordinary NVT MD simulations using a time step of 2x$10^{-15}$ sec and a Langevin thermostat with a coupling constant 1x$10^{-11}$/sec. For the strain rate of $10^3$/sec which can not be achieved through plain MD, we used our hybrid MC-MD algorithm as described in the preceeding section and in \citet{mcmd}. The time-step and thermostat were same as for the  strain rate of 5x$10^7$/sec.  All the calculations were performed on our in-house parallel MD package. Starting value of $V_0$ was picked such that it gave a rough $t_{\textbf{W}}$ of around 1 nanosecond (see table \ref{table:mcmd}). This was a value high enough for our current application. As the temperature/driving force (stress or strain) decrease, one would pick a higher $V_0$ that would accordingly lead to a larger $t_{\textbf{W}}$. During compression as work is being performed on the system, there is a change in the potential energy, and as such the value of $V_0$ was updated every few thousand MD steps by an amount equaling the change in the mean potential energy over these many MD steps. A sharp drop in the potential energy indicates that a
(partial) dislocation nucleation event has occurred in the system (see figure \ref{fig:energyprofile}). Table \ref{table:mcmd} provides values of the various parameters used in the compression testing experiment.\newline

For both of these strain rates, the yielding involves nucleation of a leading Shockley partial nucleates on a \{111\} slip plane at lower stresses than a trailing partial. This can be seen in figure \ref{fig:slip}(b) where the leading partial nucleated from the surface and left behind a 2-layer thick HCP region. The trailing and leading partials are in agreement with what one expects by calculating relative Schmid factors: for $\langle 001 \rangle$ compression, (a/6)$[11\overline{2}]$ and (a/6)$[\overline{2}11]$ were found to be the leading and trailing partials corresponding to Schmid factors of 0.47 and 0.24 respectively\citep{weinberger_jmps}. After this first nucleation event, the stress dropped down (figure \ref{fig:stressstrain}), and in all our runs it did not reach the value required for nucleating the trailing partial. Each subsequent nucleation event was found at all strain rates considered to involve another leading partial nucleation on an adjacent plane, leading thus to the formation of two twin planes (see figure \ref{fig:slip} and movie in supplementary information). With increasing deformation, the twins move apart from each other, forming a $\langle110\rangle$  reoriented wire in the middle. High strain-rate MD simulations \citep{srolovitz} have also reported the same observation, and we now provide direct evidence that the strain-rate does not affect mechanism of deformation in FCC nanowires (at least in the 5 orders of magnitude strain-rate range we considered). \newline

\begin{table}[htp]
\caption{ Starting value of $V_0$ and expected value of $t_{\textbf{W}}$, for unstrained samples at various temperatures. For strained samples, the change in potential energy during the process of straining was calculated, and $V_0$ was changed by this amount. For the temperatures between 350K to 425 K, ordinary MD was sufficient for the stress range considered in this work and hence the parameters below are only for temperatures till 325 K. }
   \centering
\begin{tabular}{c | c |c}
\hline\hline

\hline
       T(K)&
       $V_0$ (eV) &
       $t_{\textbf{W}}$ (ps) \\
       
       \hline
            275 &
       -7275.00 &
      1400 \\

       \hline
            300 &
      -7267.75 &
      1700 \\
      
           \hline
            325 &
       -7260.25 &
      30 \\
      
\end{tabular}
\label {table:mcmd}
\end{table}

Figure \ref{fig:stressstrain} does however show the quantitative effects on the yield point of unrealistic strain-rate MD calculations. We find that though the failure mechanism stays same for both the strain-rates, there is a significant difference in the strain at which slip occurs. At the high strain-rate of 5x$10^7$/sec, the wire withstands strain of as high as almost 10$\%$ before the first partial dislocation is emitted (corresponding to a stress span of around 2.6 GPa). However at the more realistic strain rate of $10^3$/sec, the partial is emitted around 8.6$\%$ only (corresponding to a stress span of around 2.35 GPa). For both the strain-rates, there is a distribution of the strain at which slip occurs and the values in \ref{fig:slip}) denote mean values.\newline

This strong difference is in accordance with the strain-rate sensitivity in true nanowires (i.e. wires less than 100nm in diameter) as predicted by \citet{juli_prl} 
and observed in real experiments on small nanowires by \citet{jenningsacta}. To understand and motivate this dependence, we look at the rate of nucleation of leading partial dislocation, as given by Eq. (\ref{eq:rate}) below \citep{juli_prl}:

\begin{equation}
\label{eq:rate}
R = N\nu(\epsilon).exp({-F(\epsilon,T)\over{k_{B} T}})
\end{equation}

Here F is the Helmholtz free energy of activation as a function of temperature T and strain $\epsilon$ (since our experimental set-up is a constant strain situation), $k_{B} T$ is the thermal energy, N is the number of equivalent surface nucleation sites and $\nu(\epsilon)$ is an athermal strain-dependent attempt frequency. Eq. (\ref{eq:rate}) thus has two contributions: an athermal part related to the elastic limit (which we defined as stress for nucleation of the first dislocation) of the surface at which nucleation would occur spontaneously without any thermal contributions, and an activated part that takes into account the role of thermal fluctuations in causing nucleation to happen even below the  athermal strain (which is the minimum strain at which nucleation would occur at absolute zero temperature).\newline

\subsection{Activation Parameters}
\subsubsection{Activation Volume}
The activation volume $\Omega$ is defined as the derivative of the activation free energy with respect to stress, i.e. $\Omega(\sigma,T) \equiv - {({\partial F\over{ \partial \sigma}})}_T $. As reported through experimental measurements as well as TST based calculations, the activation volume for surface nucleation remains in a characteristic range of few $b^3$ (where $b$ is the burgers vector). In comparison, for a typical bulk dislocation source the activation volume is upwards of 100$b^3$ and can be as high as 1000$b^3$ \citep{jenningsacta}. The activation volume in turn determines whether a process will be strain-rate sensitive or insensitive. This can be reasoned as follows. Assuming a simple case where the activation energy depends linearly on stress (see \citet{juli_prl} for detailed derivation), one can show that the most probable estimate of the nucleation stress is given by 

\begin{equation}
\label{eq:nuclstress}
\sigma = \sigma_{athermal} - {k_B T\over{\Omega}}ln{{k_B TN\nu}\over{E\dot{\epsilon}\Omega}}
\end{equation}

where E is the Young's modulus and $\sigma_{athermal}$ is the athermal nucleation stress causing instantaneous dislocation nucleation. As can be seen in Eq. (\ref{eq:nuclstress}), a high activation volume (as in the case of bulk dislocation source) masks out the effect of strain-rate. As the activation volume decreases towards values relevant for surface nucleation, the effect of strain rate should become very significant. Figure \ref{fig:stressstrain} provides the first direct MD based evidence of this. \newline

From figure \ref{fig:energy} (explained further in Section 2.4.2), we find as expected that the activation volume decreases with increase in temperature (slope of the energy versus stress profile), and that at 275 K it is around 6$b^3$. We had also calculated this quantity in \cite{mcmd} by a different method. There we re-expressed the activation volume as $\Omega = \sqrt{3}k_B T\partial({lg\dot{\epsilon}})/\partial{\sigma}$
where $\sigma$ was the stress at 11\% strain. We had then found the activation volume to be 1-2$b^3$ at 300 K, and thus the two calculations are well within order of magnitude agreement. \newline

\begin{figure}[htp]
\begin{center}
 \includegraphics[width=120mm]{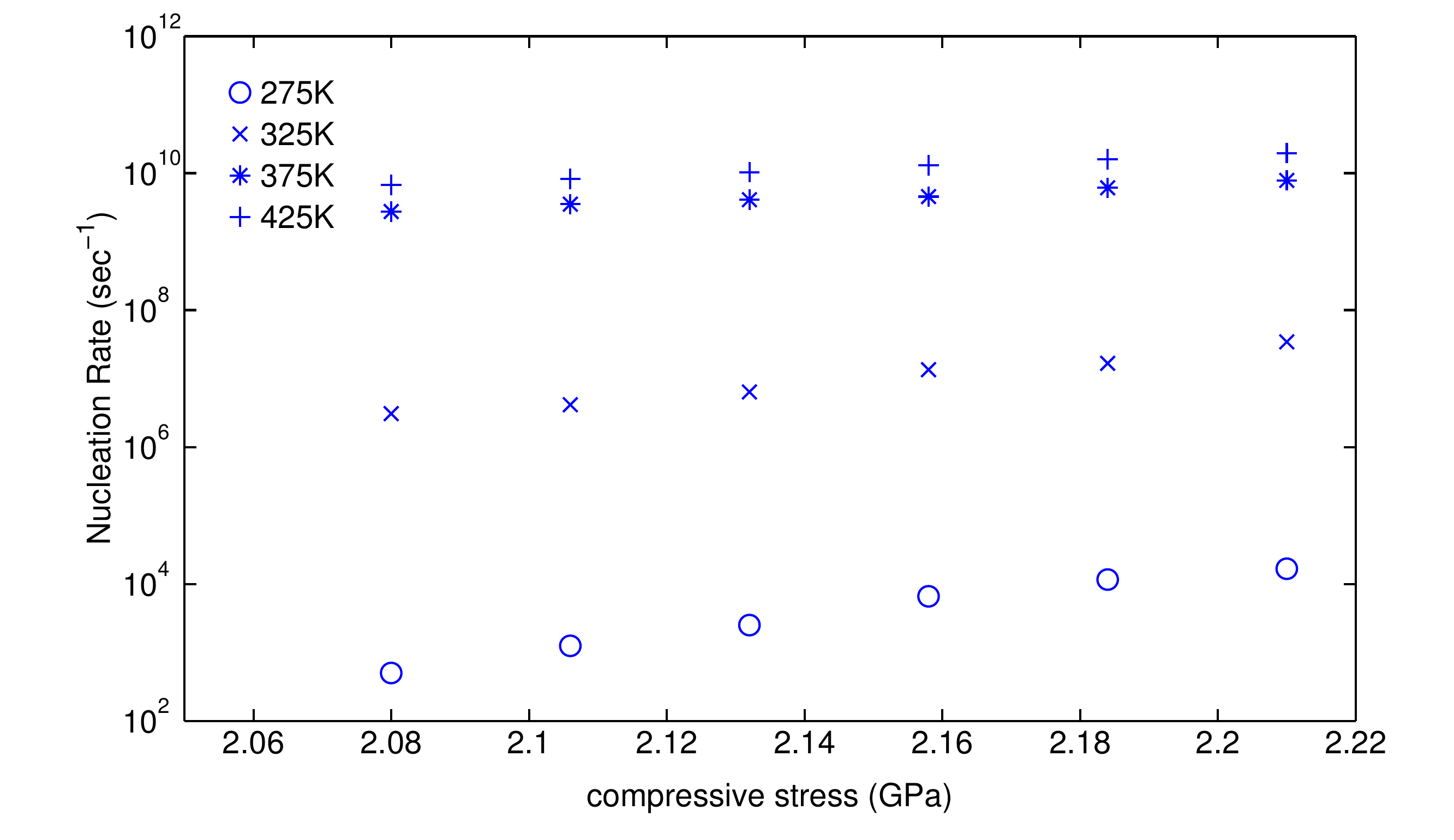}
 \end{center} 
 \caption[3]
{Strain and temperature dependence of the dislocation nucleation rate $R(\epsilon,T)$, converted here to $R(\sigma,T)$ by assuming a linear dependence of stress on strain. Each data point was calculated by averaging over 16 samples. The size of the individual markers corresponds to the 90$\%$ confidence interval in the measurement. }
\label{fig:rate}
\end{figure} 

\subsubsection{Activation Free Energy}
We now report detailed calculations of the activation terms in Eq. (\ref{eq:rate}). To do so, we performed the compression testing at a strain-rate of $10^3$/sec at 7 different temperatures from 275 K to 425 K at intervals of 25 K. These compression tests were stopped at various values of the nominal strain from 8$\%$ to 8.5$\%$ (the athermal strain at 0 Kelvin for nanowire of these dimensions was found to be around 13.5$\%$). The wisdom behind choosing this particular range of strain will be clear soon when we provide estimates of the nucleation rate. For each of these strains, the wire is still in the elastic regime. As described in the introduction, we are interested in collecting statistics of the waiting time before nucleation of the first dislocation as the nanowire is held at this strain \citep{,ngan_prl,ngan_philmag}. The structures from the compression tests stopped at varying strains served as samples for our waiting time statistics tests.\newline

For each of these structures (corresponding to a combination of imposed strain and temperature), 16 different runs were carried out where the nanowire was held at a particular strain and temperature. Each of the runs was carried out until nucleation of the first dislocation, marked by appearance of a 2-layer thick HCP region, as well as sudden dips in the potential energy and the stress (see figure \ref{fig:energyprofile}). The average rate of nucleation was then calculated as

\begin{equation}
\label{eq:avgrate}
R(\epsilon,T) = {1\over{\tau_{average}}}
\end{equation}
where $\tau_{average}$ is the time for nucleation averaged over the 16 samples. \newline

Figure \ref{fig:rate} provides the value for nucleation rate for various strain and temperature combinations. In this figure we have converted strain to stress by using the stress-strain curve in figure \ref{fig:stressstrain} for $\langle001\rangle$ compression of Gold nanowire, in order to facilitate comparison with published literature. This plot clarifies our choice of imposed strains - with 8$\%$ strain (or 2.25 GPa stress), the nucleation rate is already slower than one every few milliseconds at 275K. \cite{warner_prb} have also recently pointed out using various flavors of TST as to how the mean time for dislocation nucleation changes from picoseconds to years as the load changes by as small as 0.2 GPa. The other end of 8.5$\%$ was picked because as illustrated in figure \ref{fig:stressstrain}, the wire slips at high temperatures around 8.6$\%$. The lengthiest of these calculations took around a few CPU days. With a slightly more aggressive choice of $V_0$, it should be possible to reach the one per second or still slower regime.  \newline

For each strain $\epsilon$, we picked a sufficiently high value of reference temperature $T_{0}$ such that the rate R($\epsilon,T_0$) did not any longer depend on the choice of temperature. 
We can then make the approximation that ${F(\epsilon,T)} \gg {F(\epsilon,T_0)} $, and express Eq.(\ref{eq:rate}) as Eq.(\ref{eq:refrate}) below, to factor out the athermal frequency term. Given this approximation, the entropy of activation we calculate subsequently is effectively measured relative to the high temperature limit (since the activation entropy at high temperature could still be nonzero).

\begin{equation}
\label{eq:refrate}
 {F(\epsilon,T) } \approx -{k_{B} T} ln( {{R(\epsilon,T)}\over{R(\epsilon,T_0)}} )
\end{equation}

\begin{figure}[htp]
\begin{center}
 \includegraphics[width=110mm]{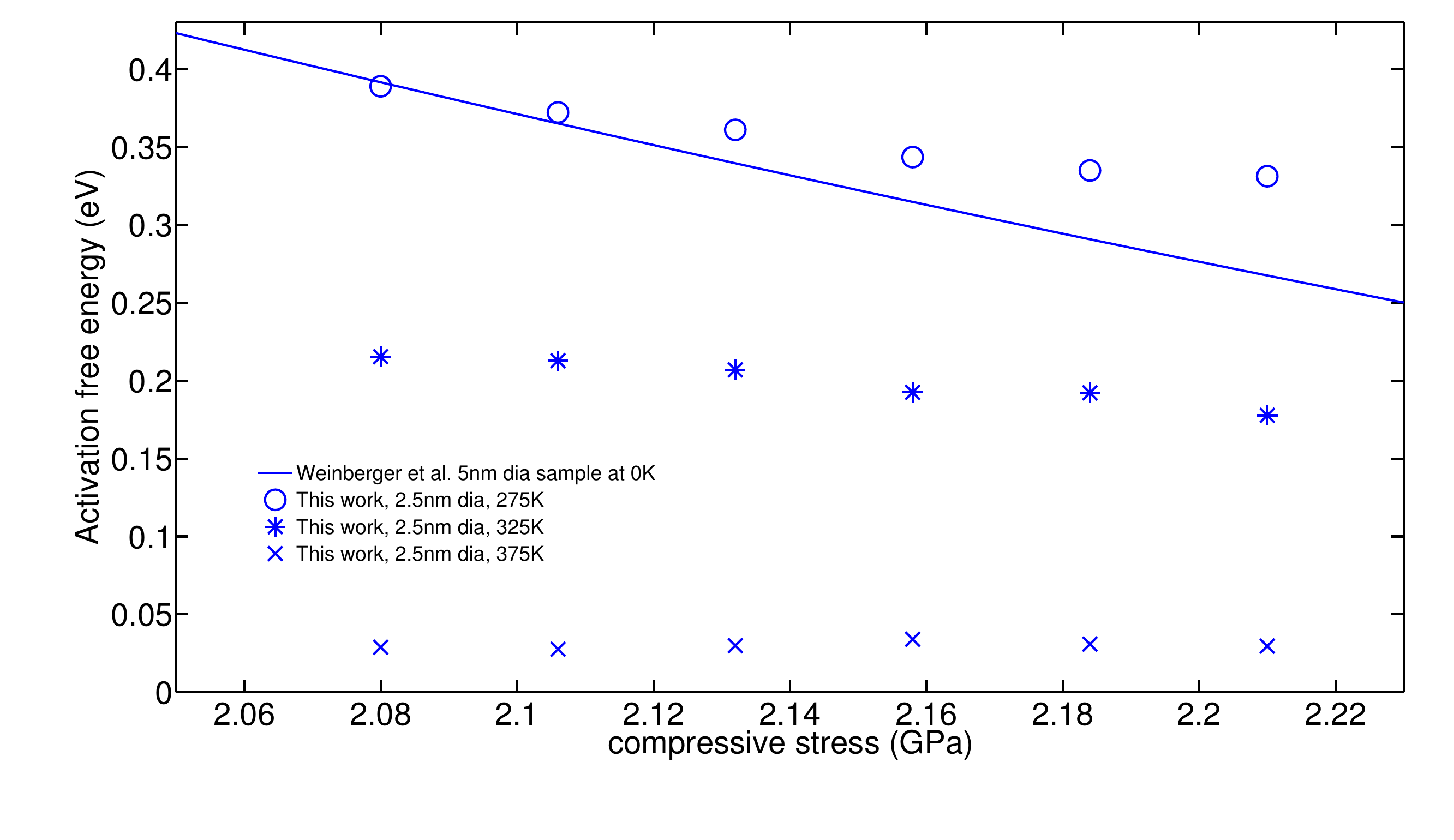}
 \end{center} 
 \caption[4]
{Activation free energy for surface nucleation of dislocations as a function of stress and temperature. Comparisons are made with results from literature for 0 Kelvin activation energy for a wire 8 times bigger (thus with many more possible nucleation sites). The size of the individual markers corresponds to the 90$\%$ confidence interval in the measurement.}
\label{fig:energy}
\end{figure}

From Eq.(\ref{eq:refrate}) we directly calculate the Activation Free Energy F($\epsilon,T$) (figure \ref{fig:energy}). Our values are in the rough benchmark of the 0.3 eV value found in Copper nano-indentation experiments where one expects homogeneous nucleation to be the mechanism at work \citep{schuh_nature}. Figure \ref{fig:energy} also provides the only published values of activation free energy at 0 Kelvin temperature for Au nanowires \citep{weinberger_jmps}.  \citet{weinberger_jmps}'s calculations are for a 5nm diameter nanowire (thus 4 times as many atoms as in our nanowire) using a chain of states methodology at 0 Kelvin. A comparison between our and their free energies is thus not really justified due to differing system sizes - this can be understood by looking at Eq. (\ref{eq:nuclstress}). In a larger sample the number of nucleation sites (surface atoms) N is higher. As such the nucleation stress goes down, increasing the probability of nucleation for the same driving force (stress/strain and temperature). This in turn leads to a lower free energy of activation. Some qualitative differences between these two calculations may will also arise because of difference in interatomic potentials. Even though a direct comparison between \citet{weinberger_jmps}'s and our results is not justified due to these reasons, we still provide their results in figure \ref{fig:energy}) since viewed together our results give a full picture of how the activation free energy varies with stress, temperature and specimen size. \newline

Our calculations demonstrate how strongly the free energy of activation depends on the temperature. It has been a common practice, mostly arising from lack of methods capable of providing high temperature activation energy barriers calculations, to assume the same temperature dependence for activation free energy across temperatures. Many workers have found direct and indirect evidence suggesting this is incorrect for studying deformation in materials. For example, \citet{curtin_actamat} found that in Al, a temperature dependent activation barrier can lead to a transition from twinning to full dislocation emission back to twinning with increasing temperature. We believe our algorithm should now provide researchers with a tool to calculate such barriers at various temperatures under realistic loading rates for the first time.

\subsubsection{Activation Entropy}
\begin{figure}[htp]
\begin{center}
 \includegraphics[width=120mm]{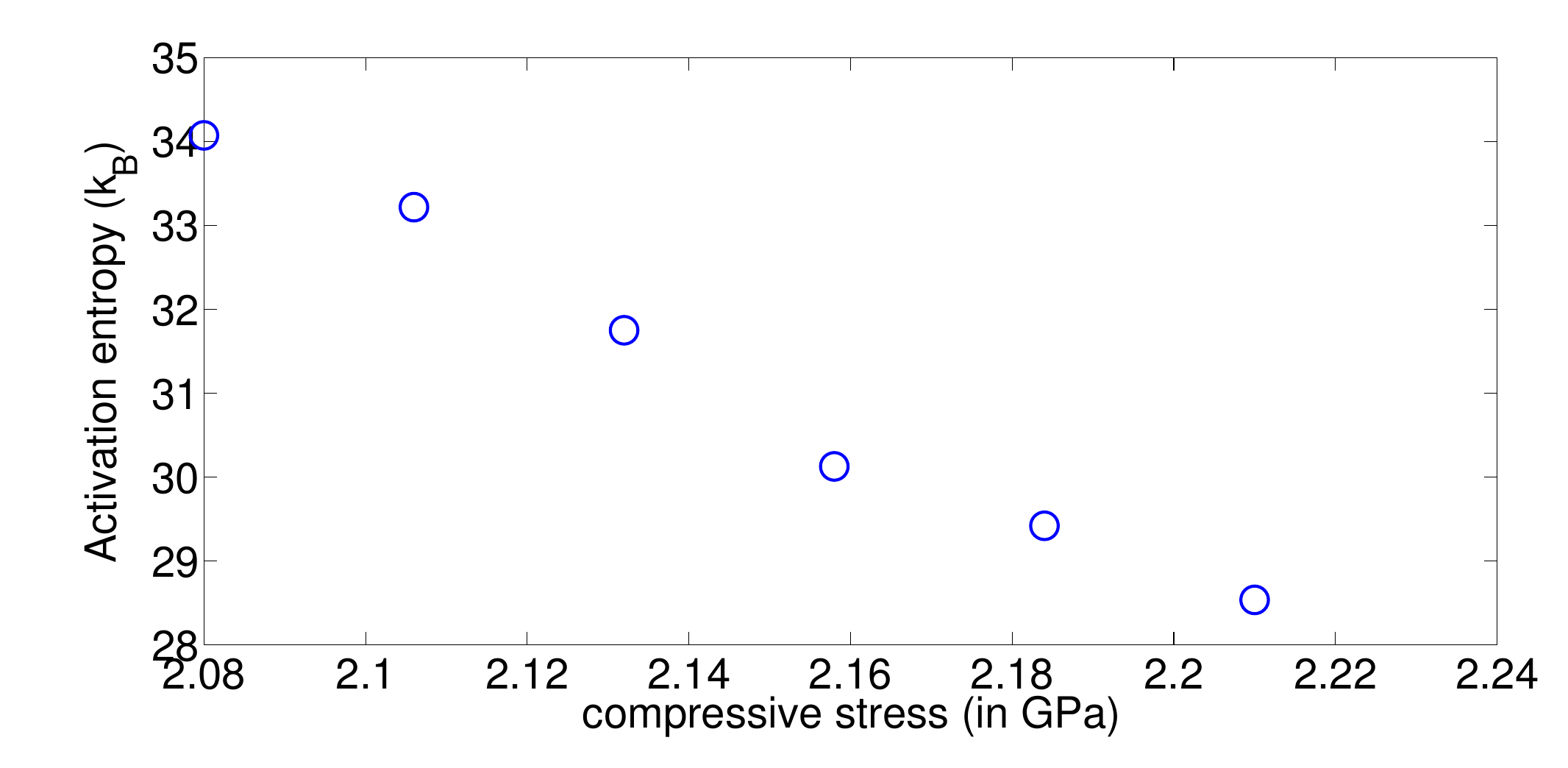}
 \end{center} 
 \caption[5]
{Activation entropy as a function of stress for surface nucleation of dislocations.}
\label{fig:entropy}
\end{figure}

From the variation of activation free energy with stress and temperature in figure \ref{fig:energy}, we calculated the stress dependent activation entropy. To obtain this quantity, we do a linear fit between the activation energy and the temperature at each stress value. The entropy is then the slope of this linear fit (with a negative sign), reported in figure \ref{fig:entropy}. Such a calculation has rarely been performed for dislocation nucleation or for other problems - the two instances of such calculations we could find were \citet{asbhd}'s recent adaptive strain-boost hyperdynamics (ASBHD) where the authors calculate the stress dependent entropy for corner nucleation in Copper nanowires, and \citet{weicai_pnas}'s Umbrella Sampling based calculations (in Copper as well). We find that the entropy decreases as the driving force (stress or strain) increases, and is typically in the range 20-30$k_B$. This is roughly in the benchmark of values reported through previous simulations. We avoid making detailed comparisons here between our values and these previously published values, given that we differ in elements (gold versus copper), geometries (circular versus square with sharp cross-sections), size and ensemble (constant stress versus constant strain, see \citet{weicai_pnas}).

\section{Discussion}
\subsection{Key results}
Our compressive tests on Au nanopillars show that the mechanism of deformation stays same for strain rates from $10^8/s$ to $10^3/s$, i.e., slip through nucleation of a leading Shockley partial dislocation on a \{111\} slip plane. However, the elastic limit (defined as stress for nucleation of the first dislocation) changes significantly as the strain-rate is changed. This is in qualitative agreement with the strain-rate dependence in nanopillars as observed in experiments \citep{jenningsacta} and also predicted by the phenomenological models of \citet{juli_prl}. We believe ours is the first fully atomistic calculation reported across 6 orders of magnitudes in the strain-rate and reaching a realistic strain rate of $10^3 /s$. \newline

The origin of activation entropy in dislocation nucleation, and related to it the rapid drop of activation free energy with temperature, can be attributed to thermal expansion in the material \citep{weicai_pnas}. As the temperature increases the expansion causes atoms to move away from each other making it easier for planes to shear and thus reducing the free energy barrier for nucleation. Our high activation entropy values show that not considering the temperature dependence of the activation energy can lead to nucleation rate being erroneous by as much as 8-12 orders of magnitude. This has been emphasized in the very recent literaure by \citet{weicai_pnas}. We also found that the entropy decreases as the driving force for nucleation (stress or strain) increases, leading to a significant dependence of the activation free energy on driving force and temperature. Our method now provides an easy-to-implement way to calculate this dependence under realistic driving forces for any general sample geometry.\newline

\subsection{Comparison of our algorithm with other algorithms for extended time-scales}
It is also instructive to compare our algorithm with other realistic time-scale algorithms for studying dislocation nucleation.
Two algorithms that have been recently used to model temperature-dependence of the dislocation nucleation process include the Umbrella Sampling method (proposed by \citet{weicai_pnas}) and the Adaptive Strain Boost Hyperdynamics (ASBHD) method (proposed by \citet{asbhd}). Both are excellent methods, but have their respective limitations. \newline

The former needs to define an order parameter (or a reaction co-ordinate) for the system, in terms of which the biasing potential is then imposed on a certain sub-set of atoms. Picking such an order parameter might be a non-trivial task in complicated sample geometries and can have its pitfalls \citep{rarepitfalls}. Also, while this method is excellent for computing free energy barriers - it does not perform any actual dynamics. ASBHD does away with the need to pick up an order parameter. However the speed-up in ASBHD relative to ordinary MD becomes significant (i.e. 4-5 orders of magnitude or more) only as the temperature of the system falls below 100K or so. The method is a local boost scheme, thus specific atoms are lifted out of the low-lying energy basins, making them preferential sites for nucleation to happen. Such a local boost scheme works well for specific geometries (such as a square nano-rod with sharp corners), but might not be very well suited for studying more homogeneous nucleation as we are considering in the current paper. Since our method does not require the specification of the degrees of freedom of interest (which is crucial when the mechanisms are complex and involve the movement of many atoms), it is well suited for studying homogeneous nucleation. \newline

By providing a time-scale correction independent to the main simulation, our algorithm also provides the ability to implement time-dependent boundary conditions (relevant to a tensile test for e.g.) and in general time-dependent forces. To do this, we can launch a set of adiabatic switching jobs prior to the main simulation, that give a good starting value for the quantity $t_{\textbf{W}}$ and thus for the boosted time-scale. This is to be contrasted with ASBHD \citep{asbhd} and Hyperdynamics methods in general \citep{voter_prl} where time-scale estimates remain noisy and non-converged for long simulation times, especially as one tries to increase the speed-up relative to ordinary MD \citep{fichthorn,mcmd}. \newline

Thus, the main advantage of our method over these previous methods used for studying dislocation nucleation is that we can sample wells and transition states without any prior knowledge or without favoring any atoms for being the nucleation sites. Thus our method can be expected to perform well at capturing the dynamics of events beyond the first nucleation event, irrespective of sample geometry. \newline

\subsection{Further discussions our algorithm}
One obvious limitation of our method is the choice of the parameter $V_0$. For high speed-ups one would like $V_0$ to be high. But caution needs to be exercised since a high $V_0$ can cause distinct wells to coalesce into a single well, thus coarsening our description of the dynamics. Another limitation is that, as detailed in Section 2.2.1, we allow MD to run for some time before concluding that the system is now inside the well. This helps us deal with the re-crossing trajectory problem that makes TST based expressions incorrect \citep{voter_review}. The longer one allows MD to run before calling the next instance of MC, the more accurate does the boosted time-scale become. This can however lead to the speed-up relative to ordinary MD going down.\newline

The proposed method is correct for any choice of V$_0$ in the sense that the calculated escape rate from a basin defined by an upper potential energy threshold V$_0$ (in which the system has equilibrated) is correct. It may be that, in some systems, the shape of a basin defined by an upper potential energy threshold V$_0$ is complex and makes it difficult to interpret the results and/or slows down the equilibration of the system within a well. The method may be less useful or less efficient in such cases, but not incorrect. We describe below that in what kind of systems would this problem be especially serious and when it would not:
\begin{itemize}
\item{	The worst-case scenario is when the system consists of a few specific atoms performing “interesting” dynamics embedded in a large number of other atoms experiencing nothing but “uninteresting” oscillations. In that case, it would be difficult to delineate the basins corresponding to the “interesting” atom hops because of the large thermal noise of the “uninteresting” atoms. In this situation, the basins may be connected by a few very “thin” tunnels that are rarely taken, not because of a high energy barrier, but because there are so few of them. Our method would not improve the speed very much in that case because it only boosts the rate of events with high barrier. However, this would be precisely the type of system where it would be easy to construct suitable collective coordinates, reaction coordinates or identify bonds \citep{metadynamics,fichthorn,asbhd} to be boosted, so other accelerated methods could be easily used.}
\item{	If the number of candidate “interesting” events grows with system size, then it is difficult to identify the “interesting” degrees of freedom \textit{a priori} and a method such as ours become very useful. In addition, since there are many possible paths that the system can take out of a basin in this case, the only reason for these events being rare are their high energy barrier. In such a situation, our method performs best. Coincidentally, our dislocation nucleation problem is one where the number of candidate events does grow with system size because there are many possible dislocation nucleation sites due to the cylindrical geometry.}
\end{itemize}

 The recently proposed $\kappa$-dynamics method by \cite{kappa} bears some similarity to our approach in terms of accelerating transition from one well to another, and carefully correcting for trajectories that recross the transition surface and re-enter the same well. Their method however requires (a) requires significant computer resource to compute the transition surface and (b) identification of a good reaction co-ordinate. Our approach differs from $\kappa$-dynamics in how it deals with re-crossing events. In $\kappa$-dynamics, multiple attemps to exit the well are made until one is found that does not recross the transition surface. In our method, we handle this issue at the entrance of the well, by waiting until the system has spend a sufficient amount of time in a well (so that it is equilibrated) before calculating the flux towards the outside of the well (this is more similar to what is done in the parallel replica method \citep{parrep,parrep_driven}). We also provide an efficient way to compute this flux via Monte-Carlo and adiabatic switiching that avoids noisy averages of terms that are exponential in the total energy which would converge slowly. In contrast, $\kappa$-dynamics relies on umbrella sampling, which involves averages of terms that are exponential in the total energy. It might be possible to combine some of the good features of both $\kappa$-dynamics and our method to come up with a yet more robust method.

\section{Conclusions}
In this paper we have derived and demonstrated a hybrid MC-MD algorithm that can be used to achieve realistic time-scales in fully atomistic simulations of materials while still predicting correct deformation physics. The algorithm is especially designed to be suited for massive parallelization. By using this algorithm, we obtained compression testing stress-strain plots at strain rates several orders of magnitude lower than ever previously reported for MD simulations. We showed that high strain-rates in simulations, which have been common due to lack of methods capable of implementing low strain-rates, can lead to a significant error in the elastic limit (defined earlier) of the material. We also derived the full stress and temperature dependence of the activation free energy for surface nucleation of dislocations in Gold nanowires. The algorithm was described in sufficient detail to be useful to the mechanics community for different applications. \newline

\textbf{Acknowledgments}

We would like to thank Dr. Andrew Jennings and Prof. Julia Greer for helpful discussions and comments on the manuscript; Dr. Arthur Voter for helpful discussions regarding the algorithm; and Dr. Seunghwa Ryu for originally suggesting this particular application at a Gordon Research Conference. This research was supported by the US Office of Naval Research via grant N00014-12-1-0557, US National Science Foundation under CAREER Grant DMR-1154895 and by XSEDE computational Resources provided by NCSA and TACC under Grant No. DMR050013N. \newline

\small{
\bibliographystyle{elsarticle-harv}
\bibliography{jmps}
}






\end{document}